\documentclass{bioinfo}
\copyrightyear{2015} \pubyear{2015}

\access{Advance Access Publication Date: Day Month Year}
\appnotes{Manuscript Category}
\usepackage{booktabs}
\usepackage{float}
\usepackage{graphicx}

\usepackage[section]{placeins}
\usepackage{stfloats}
\usepackage{natbib}
\usepackage{cite}
\usepackage{caption}

\begin{document}

\firstpage{1}

\subtitle{Subject Section}

\title[Graph2MDA]{Graph2MDA: a multi-modal variational graph embedding model for predicting microbe-drug associations}
\author[\textit{et~al}.]{Lei Deng\,$^{\text{\sfb 1}}$, Yibiao Huang\,$^{\text{\sfb 1}}$, Xuejun Liu\,$^{\text{\sfb 2,*}}$, Hui Liu\,$^{\text{\sfb 2,*}}$}
\address{$^{\text{\sf 1}}$School of Computer Science and Engineering, Central South University, Changsha, 410083, China. \\
$^{\text{\sf 2}}$School of Computer Science and Technology, Nanjing Tech University, 211816, Nanjing, China}

\corresp{$^\ast$To whom correspondence should be addressed.}

\history{Received on XXXXX; revised on XXXXX; accepted on XXXXX}

\editor{Associate Editor: XXXXXXX}

\abstract{
\textbf{Motivation:} Accumulated clinical studies show that microbes living in humans interact closely with human hosts, and get involved in modulating drug efficacy and drug toxicity.  Microbes have become novel targets for the development of antibacterial agents. Therefore, screening of microbe-drug associations can benefit greatly drug research and development. With the increase of microbial genomic and pharmacological datasets, we are greatly motivated to develop an effective computational method to identify new microbe-drug associations.\\
\textbf{Results:} In this paper, we proposed a novel method, Graph2MDA, to predict microbe-drug associations by using variational graph autoencoder (VGAE). We constructed multi-modal attributed graphs based on multiple features of microbes and drugs, such as molecular structures, microbe genetic sequences, and function annotations. Taking as input the multi-modal attribute graphs, VGAE was trained to learn the informative and interpretable latent representations of each node and the whole graph, and then a deep neural network classifier was used to predict microbe-drug associations. The hyperparameter analysis and model ablation studies showed the sensitivity and robustness of our model. We evaluated our method on three independent datasets and the experimental results showed that our proposed method outperformed six existing state-of-the-art methods.  We also explored the meaningness of the learned latent representations of drugs and found that the drugs show obvious clustering patterns that are significantly consistent with drug ATC classification. Moreover, we conducted case studies on two microbes and two drugs and found 75\%-95\% predicted associations have been reported in PubMed literature. Our extensive performance evaluations validated the effectiveness of our proposed method.\\
\textbf{Availability:} Source codes and preprocessed data are available at  https://github.com/moen-hyb/Graph2MDA
\\
\textbf{Contact:} \href{hliu@njtech.edu.cn}{hliu@njtech.edu.cn}\\
\textbf{Supplementary information:} Supplementary data are available at \textit{Bioinformatics}
online.}

\maketitle

\section{Introduction}
Microbe communities, including bacteria, archaea, viruses, protozoa, and fungi, are closely associated with human hosts (\citealp{0Structure, 2013The}). They reside in various human organs, such as skin, gastrointestinal tract, oral cavity, and other tissues. Accumulated studies have confirmed that microbe plays a fundamental role in keeping the homeostasis of the human internal
environment. Their beneficial functions cover quite a few aspects, such as improvement of metabolism, synthesis of essential vitamins, resistance to pathogens, and enhancement of immunity (\citealp{2009Genome}). For example, gut microbes play a key role in modulating immune response or immune tolerance through Treg cell regulation(\citealp{2018Exploring}). Also, the microbiome acts in collaboration with host mucosal sites as a barrier to pathogens (\citealp{2004Interactions}).

As such, imbalance of microbe communities often leads to various diseases, such as diabetes (\citealp{2008Innate}), obesity (\citealp{2010Human}), and even cancer (\citealp{2013cancer}). Also, several studies have shown that microbes involve in drug absorption and metabolism, and thus modulate the drug efficacy and drug toxicity (\citealp{2021Towards}). For instance, the gut Actinobacterium \textit{Eggerthella lenta} is responsible for the inactivation of the cardiac drug \textit{digoxin} (\citealp{haiser2013predicting}). The vaginal microbiota is effective at degrading \textit{tenofovir} before the host cell converts the drug into its drug active form (\citealp{2017Vaginal}). On the other hand, drugs also, in turn, change the diversity and function of microbe communities living in the human body. Some new drugs have been developed to target the microbes to maintain the fitness of microbial community structure (\citealp{2017Microbiome}). But everything has two sides, the interplay between microbes and drugs leads to the spread of drug-resistant bacteria, which poses another serious threat to human health (\citealp{2014Antimicrobial}). As a result, there is an urgent need to identify the associations between microbes and drugs. 

As conventional wet-lab assay is time-consuming and labor-intensive, \textit{in silico} methods have been proposed to prioritize microbe-drug associations for further experimental validation. \textit{HMDAKATZ} (\citealp{chen2017novel}) adopted KATZ metric to prediction microbe-drug associations, while \textit{NTSHMDA} (\citealp{luo2018ntshmda}) used random walk with restart on microbe-disease heterogeneous network to predict new associations. Recently, several databases have released a large number of experimentally validated associations of microbes with drugs, such as MADA (\citealp{sun2018mdad}), Abiofilm (\citealp{2017aBiofilm}) and Drugvirus (\citealp{andersen2020discovery}), as well as the COVID-19 database HDVD (\citealp{meng2021drug}).  These freely accessible datasets motivated the development of deep learning-based methods to predict microbe-drug associations. For example, \textit{EGATMDA} (\citealp{long2020ensembling}) constructed a novel integrated graph attention network to predict human microbe-drug associations. \textit{GCNMDA} (\citealp{long2020predicting}) engineered conditional random field into the graph convolutional network framework to predict microbe-drug association. \textit{LAGCN} (\citealp{yu2020predicting}) is another computational model based on layered attention map convolutional network. The studies demonstrated their progress on the predictive performance compared to traditional classifiers, but they still have limitations that seriously restrict their effectiveness. These models constructed heterogeneous networks and extract features to predict new associations. However, they failed to tackle outlier nodes in the feature extraction, which often resulted in incorrect predictions. Also, these models take into account either network structural information or node features, but failed to deal with both simultaneously.

In this paper, we proposed a novel model, Graph2MDA, which is an integrated framework of Variational Graph Autoencoder (VGAE) and Deep Neural Network (DNN) to prioritize microbe-drug association. To make full use of multiple types of attributes that can be derived from microbes and drugs, we constructed multi-modal attribute graphs composed of microbes and drugs and their associations. For drugs, the molecular structure, drug interaction profile, and network topological attributes were taken into account. For microbe, the genomic sequence and functional annotations are considered. Based on the multi-modal attribute graphs, we trained VGAE to learn informative and interpretable latent representations of each node and whole graph.  Next, a DNN classifier is fed by the learned latent representation and outputs the probability standing for the presence of microbe-drug associations. We have performed extensive performance evaluations and demonstrated that Graph2MDA can effectively leverage multi-modal attribute networks to improve its performance. On other independent datasets, Graph2MDA outperformed the other six state-of-the-art methods. Especially, we found that the drugs show obvious clustering patterns in the latent representation space, and the clusters are significantly consistent with drug ATC classification. This validated the interpretability of the learned latent representations. Finally, our case studies on two microbes (i.e. human immunodeficiency virus, Mycobacterium tuberculosis) and two drugs (i.e. Cloxacillin and Aloe Vera Gel) also proved the effectiveness and robustness of our proposed model. To our best knowledge, we are the first to apply VGAE to the prediction of microbe-drug associations.

\section{Materials and methods}
\subsection{Data sources}
The experimentally confirmed microbe-drug associations are obtained from the MDAD database(\citealp{sun2018mdad}). The database has collected a total of 2,470 clinical reports or experimentally verified associations between 1,373 unique drugs and 173 kinds of microbes.

The drug-drug interactions are retrieved from the Drugbank database (\citealp{wishart2018drugbank}). We selected the interactions associated with the drugs included in the MDAD dataset and got 5,586 drug-drug interactions covering 1,228 drugs. The microbe-microbe interactions are derived from the MIND database~\footnote{http://www.microbialnet.org/mind\_home.html}. Similarly, we selected only the interactions associated with the microbes included in the MDAD dataset and got 138 microbe-microbe interactions covering 123 microbes. The FASTA genome sequence of microbes is downloaded from the NCBI database, and 131 of the 173 microbes are found. The detailed numbers of the aforementioned datasets are shown in Table~\ref{Tab:01}.

\vspace{-0.5 cm}
\begin{table}[h]
	\processtable{The details for each microbe-drug network. \label{Tab:01}}
	{\begin{tabular}{p{3.5cm}p{1cm}p{1cm}p{1.5cm}}\toprule
			Networks & microbes & drugs & associations\\\midrule
			Bipartite network & 173 & 1373 & 2470\\
			drug-drug  network & ---- & 1228 & 5586 \\
			microbe-microbe  network & 123 & ---- & 138 \\\botrule
	\end{tabular}}{}
\end{table}

\subsection{Multi-modal attribute graph construction}
Given the data source of drugs and microbes, we built the bipartite network (BipNet) using only the microbe-drug associations. Specifically, the node-set of the BipNet network includes only the microbes and drugs involved in known associations, as shown in Figure~\ref{fig:01}. Also, we constructed the heterogeneous network (HetNet) that includes microbe-drug association, as well as the drug-drug and microbe-microbe associations. Note that the HetNet network contains additional nodes that are not appear in the BipNet network. Without loss of generality, denote by  $A \in\ R^{nd\times nm}$ the adjacent matrix of the network, in which $nd$ and $nm$ represent the number of microbes and drugs, respectively. If there exists a known association between node $i$ and $j$, element $A_{ij}$ is 1, and otherwise 0.

\subsubsection{Drug similarity attribute}
\paragraph{\textbf{Molecular structure similarity }}
We used SIMCOMP2 (\citealp{hattori2010simcomp}) tool to calculate drug structural similarity. SIMCOMP2 measures the similarity between drugs based on their chemical structure information. We computed the pairwise drug structural similarity and then constrcted the similarity matrix ${DS}^{struct}\in R^{\mathbf{n}d\times\mathbf{n}d}$, where $ {DS}^{struct}(d_i,d_j) $ represents the molecular structural similarity between drug $d_i$ and drug $d_j$.  

\paragraph{\textbf{Drug gaussian kernel similarity}}
With the assumption that drugs with similar therapeutic effects interact closely with similar microbes, we used the Gaussian kernel interaction of drugs to calculated another similarity measure. Denote by $DIP(d_i)$ the drug-drug interaction profile of drug $d_i$, namely the $i$-th row of drug-drug interaction matrix representing the interactions between drug $d_i$ and all other drugs,
the Gaussian kernel similarity ${DS}^{gauss}(d_i,d_j)$ between drug $d_i$ and drug $d_j$ is formulated as below:
\begin{equation}
  DS^{gauss}(d_i,d_j)=exp(-\mu {\parallel\!DIP(d_i)-DIP(d_i)\!\parallel}^2)
\end{equation}
in which $\mu$ represents the normalized kernel bandwidth, defined as follows:
\begin{equation}
 	\mu=\mu^\prime{(\frac{1}{nd} \sum^{nd}_{i=1}{\parallel\!DIP(d_i)\!\parallel}^2)}^{-1}
\end{equation}
Where $\mu^\prime$ is the original bandwidth and is generally set to 1.

\paragraph{\textbf{Fused drug similarity}}
We combined the molecular structure similarity ${DS}^{struct}(d_i,d_j)$ and the drug Gaussian kernel similarity ${DS}^{gauss}(d_i,d_j)$ to get the integrated drug similarity $S_d(d_i,d_j)$, and the new drug similarity is calculated as follows:

\begin{equation}
	S_d(d_i,d_j) = \frac{DS^{struct}(d_i,d_j)+DS^{gauss}(d_i,d_j)}{2}
\end{equation}

\subsubsection{Drug network topological attribute}
Since several drugs are far from the majority in terms of the similarity measure, we ran random walk with restart (RWR) on the drug-drug network to derive another drug attribute. RWR can effectively capture the inherent features of the local and global topological information of a network and has been widely used in image recognition to reduce noise(\citealp{jain2018random}). Formally, RWR is defined as:
\begin{equation}
	p_i^{(t+1)}=(1-\theta)p_i^{(t)} T + \theta p^{(0)}_i
\end{equation}
in which $\theta$ is the restart probability, and $T$ is the transition probability matrix, $ p^{(0)}_i\in R^{n\times1}$ represents the starting probability vector of the i-th node, and $p_i^{(t)}\in R^{n\times1}$ denotes the probability of node $i$ moving to other nodes at time $t$. After convergence of RWR, we obtain the probability distribution vector of each drug. As such, the probability distribution vectors are used as the network topological attribute matrix $F_d\in R^{nd \times nd}$. With the integration of structural feature, the attribute of outlier nodes is effectively enriched, and lead to informative and interpretable latent representations.

\subsubsection{Microbe functional similarity}
We used the Kamneva tool (\citealp{kamneva2017genome}) to calculate the microbe functional similarity. To obtain the functional similarity between the two microbes, a microbial protein-protein functional association network was constructed, in which the nodes represent any gene family encoded by the genome, and the link represents a genetic neighbor score based on the latest STRING database(\citealp{2020The}). Then, the functional similarity of the microbe was calculated as the ratio of the link scores connecting the two microbes to the sum of all the link scores of the two microbial gene families. Finally, a matrix $S_m\in R^{nm\times nm}$ is used to represent the microbe function similarity, where $S_m(m_i,m_j)$ represents the similarity between microbe $m_{i}$ and microbe $m_{j}$ .

\subsubsection{Microbe sequence attribute}
The original genome sequences were encoded by one-hot coding, and all sequences were filled with 0 to make the length of all sequences the same. For microbes with no sequence found in NCBI, we use the mean of other known microbes instead. Finally, we ran principal component analysis (PCA) to extract the main dimensional feature of microbes (\citealp{chen2002principle}). The microbe sequence attribute is expressed by the $k$-dimensional matrix as $F_m\in R^{ nm\times k}$.

\subsubsection{Mutli-modal attribute construction}
To make the descriptive values of microbes (drugs) comparable across multiple types of measures, we separately normalized the similarity-based attribute matrix $S_d$ and topology-based attribute matrix $F_d$ for drugs, as well as the $S_m$ and $F_m$ for microbes. Then, we constructed multi-modal attributes for microbes and drugs. In fact, we could construct some modals by combining the different attributes above. For simplicity, we consider only two different modals. The first is merely based on the similarity measure, denoted by $X_{similarity}\in R^{(nd+nm)\times(nd+nm)}$ and defined as below:
\begin{equation}
	X_{similarity}\ =\ \left[\begin{matrix}0&S_m\\S_d&0\\\end{matrix}\right]
\end{equation}

The other modal integrated the secondary features extracted from the drug-drug interaction network, and the main component of microbe sequence. So, it is referred to as secondary attribute, denoted by $X_ {secondary}\in R^{(nd+nm)\times(nd+k)}$:
\begin{equation}
	X_{secondary}\ =\ \left[\begin{matrix}0&F_m\\F_d&0\\\end{matrix}\right]
\end{equation}
The two different attributes of drugs and microbes are separately or combinedly taken as input into the VGAE learning framework, and we evaluate their influence on the predictive performance in our experiments.

\subsection{Our learning framework}
We proposed a novel model, Graph2MDA, an integrated framework based on Variational Graph Autoencoders (VGAE) and Deep Neural Networks (DNN), to predict the association between microbes and drugs in multi-mode networks. As shown in Figure 1, VGAE is used to learn the latent representations (embedding) by taking as input the graph and node attributes constructed from raw data. The deep neural network (DNN) classifier receives the learned embedding by the VGAEs to predict microbe-drug associations. Note that we used two VGAE when BipNet and HetNet are simultaneously feed, and the output embedding vectors are concatenated to form the input of the DNN classifier.

\subsubsection{Variational graph autoencoder}
Variational Graph Autoencoder(VGAE) (\citealp{kipf2016variational}) is a an unsupervised learning framework based on variational autoencoder (VAE). VAE (\citealp{kingma2013auto}) is a deep generative model in the combination of variational Bayesian inference and neural networks, and it provides a generic probability modeling for describing a variable in latent space. VGAE extends VAE to learn interpretable latent representations of graph-structured data. VGAE uses a graph convolutional network (GCN) as the backbone of the encoder to learn the latent node-level and graph-level representations, which is used by the decoder to reconstruct the graph. VGAE has made great achievements in many bioinformatics tasks, such as miRNA-disease association prediction (\citealp{ding2020variational}), lncRNA-disease association prediction (\citealp{shi2021representation}). However, VGAE has not been applied to predict microbe-drug associations, and current studies did not make full use of node similarity and topological attributes.

\subsubsection{Graph2MDA model}
We defined an undirected graph $G$=($V, E$), where $V$=$\{v^{(m)},v^{(d)}\}$ represents the set of microbe and drug nodes, and each edge $e_{ij}\in E$ represent an association between a pair of different type nodes (microbe and drug) or same type nodes (two microbes or two drugs). Denote the adjacency matrix of $G$ by $A$, and the attribute matrix by $X$ served by $X_{similarity}$ or $X_{secondary}$ or a combined form of both.

\begin{figure*}[htb]
	\centerline{\includegraphics[width=\textwidth]{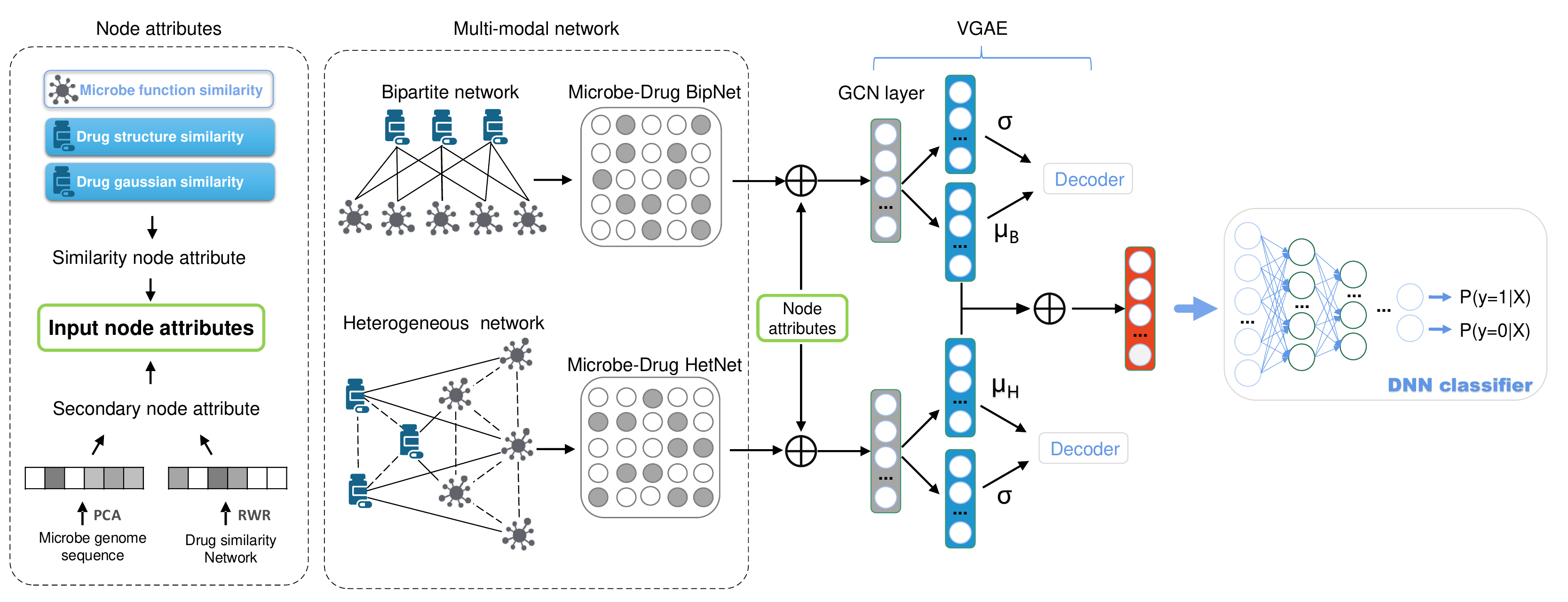}}
	\caption{Illstrative architecture of Graph2MDA learning framework for microbe-drug association prediction. Based on the multi-modal attribute graphs constructed from multiple type of attributes of microbes and drugs, two VGAE models are trained to learn informative and interpretable latent representations for node attributes and whole graphs. Next, a DNN classifier is fed by the learned latent representations to predict microbe-drug associations.}
	\label{fig:01}
\end{figure*}

As shown in Figure \ref{fig:01}, the GCN encoder converts each node $v_i$ in the graph to a low-dimensional latent representation by aggregating the informations from its local neighborhood nodes and the attributes itself. Next, the decoder tries to reconstruct the adjacency relationship $A_{ij}$ corresponding to the node pair $v_i$ and $v_j$. The encoder and decoder are trained to optimize the latent representations by progressive message propagation among the microbes and drugs in the heterogeneous network. Finally, the learned latent representations are fed into a DNN classifier to predict new microbe-drug associations. 

\paragraph{Encoder}
The encoder is a two-layer graph convolutional network (GCN) (\citealp{kipf2016semi}), which takes the network adjacency matrix $A$ and the node attribute matrix $X$ as input and output the latent variable $Z$. More specifically, we tried to model the conditional probability distribution $q(Z|X, A)$ using Normal distribution by a two-layer GCN as below:
\begin{equation}
	q(Z|X,A)=N(Z;\mu,\sigma^2I)
\end{equation}
in which $\mu$ and $\sigma$ are the mean and variance of the Gaussian distribution with respect to the latent variable $Z$, respectively. $I$ is an identity matrix. The two-layer GCN is defined as follows:
\begin{equation}
	\bar{X}=GCN(X,A)=\widetilde{A}ReLU(\widetilde{A}XW^0)W^1,
\end{equation}
in which
\begin{equation}
	\widetilde{A}=D^{-\frac{1}{2}}AD^{-\frac{1}{2}}
\end{equation}
and $W^i$ represents the parameters of $i$-th GCN layer we need to train, $ReLU=max(0,\bullet)$ is an element-wise activation function, $D$ is the degree matrix of $A$, and $\widetilde{A}$ is a symmetric normalized adjacency matrix. $A$ is first normalized by symmetric normalization to maintain the size of the eigenvector so that the sum of all rows is 1. With the GCN output, we  calculated the mean and variance of the Gaussian distribution as:
\begin{equation}
	\mu={\rm GCN}_\mu(\bar{X},A)=\widetilde{A}\bar{X}W_\mu
\end{equation}
\begin{equation}
	log\sigma={\rm GCN}_\sigma(\bar{X},A)=\widetilde{A}\bar{X}W_\sigma
\end{equation}
where $W_\mu$ and $W_\sigma$ represent the parameters of layer $\mu$ and layer $\sigma$  we need to train. Once we obtain the value of $\mu$ and $\sigma$, $Z$ can be sampled from the distribution $q(Z|X,A)$ using the reparameterization technique (\citealp{kingma2013auto}), that is, $z_i$ can be calculated by the following formula:
\begin{equation}
	z_i=\mu+\sigma\ast\varepsilon_i
\end{equation}
Where ${\varepsilon_i} \sim N(0,1)$.

Intuitively, the GCN encoder progressively aggregates information from neighbors (including both microbes and drugs) to update the attribute for each node, so that the latent variable of each node (microbe or drug) yields to an informative representation that integrates both structural and attributes information. This is particularly useful for microbes (or drugs) with few annotations. 

\paragraph{Decoder}
Considering that the latent variables $Z$ learned by the encoder already carry sufficient information, the decoder runs a simple inner product to reconstruct the adjacency matrix $A$. Formally, let ${p(A_{ij}|z_i,z_j)}$ be the conditional probability having an edge between node $i$ and $j$ given latent variable $z_i$ and $z_j$, we have
\begin{equation}
	p\left(A\middle|Z\right)=\prod_{i=1}^{N}\prod_{j=1}^{N}{p(A_{ij}|z_i,z_j)},
\end{equation}
in which $p\left(A_{ij}\middle| z_i,z_j\right)$ is defined as
\begin{equation}
	p\left(A_{ij}\middle| z_i,z_j\right)=\varphi({z_i}^Tz_j),
\end{equation}
where $\varphi(\bullet)$ is a sigmoid function. The sigmoid function we use transforms take as input the inner products to estimate the probability of associations between microbes and drugs. The decoder is trained to output $\hat{A}$ close to the real adjacency matrix $A$.

\paragraph{Loss function}
The loss function consists of two parts, the first part $E(*)$ is the binary cross entropy between input graph $A$ and output graph $\hat{A}$, and the second part is the KL divergence between $q(Z|X,A)$ and $p(A|Z)$:
\begin{equation}
	L=E_{q(Z|X,A)}\left[\log p\left(A\middle| Z\right)\right]-KL[q(Z|X,A)||p(Z)]
\end{equation}
According to the setting in VAE, we assume $p(Z) \sim N(0,1)$. During training, we used stochastic gradient descent to train VGAE to minimize the loss function.

\subsubsection{DNN classifier}
We formulated the prediction of the association between microbes and drugs into a binary classification task. A multi-layer fully connected DNN is used as the classification model. The DNN classifier takes as input the latent representations of microbes and drugs, transforms the information through multiple non-linear hidden layers, and outputs the probability standing for the existence of the association between two nodes. The cross entropy is used as the loss function, which is first pre-trained using the adaptive optimizer Adam, and then fine-tuned using the Stochastic Gradient Descent (SGD) optimizer. As the performance of the DNN classifier is affected by the hyperparameters, we empirically evaluated the impact of hyperparameters on the performance shown in detail in the subsequent experiments.According to the setting in VAE, we assume $p(Z) \sim N(0,1)$. During training, the stochastic gradient descent algorithm was used to train VGAE to minimize the loss function. 

\section{Results}

\subsection{Hyperparameter sensitivity analysis}
There are several hyperparameters in the VGAE and DNN classifier. To explore the impact of the hyperparameters, we performed a 10-fold CV on the MDAD dataset to observe the changing trend of performance so as to tune their values.

We first investigated the effect of the number of hidden layers of the VGAE encoder and DNN classifier. As too many hidden layers lead to overfitting and meaningless latent representations, we considered most 4 hidden layers in the VGAE encoder and DNN classifier. We adopt a tower-shaped DNN structure, as suggested by He et al. that the tower-shaped DNN model can extract more compact and discriminative features at a higher level (\citealp{he2016deep}). We tested the number of hidden layers in \{1,2,3,4\}, and the size of the hidden layer is set to 1024, 512, 256, 128, respectively. As shown in Figure 4 (a), the AUC value increases when the number of hidden layers of the DNN classifier increases, achieves the best performance at 3, and decreased once the layer number exceeds 3. Thus, we choose a three-layer tower structure for the DNN classifier model.  As shown in figure 4 (b), when the number of VGAE hidden layers is 2, the performance is the highest and drops sharply once the layer number exceeds 2.

Moreover, we explored the impact of dropout and learning rate (lr) in the VGAE encoder. As shown in Figures 4 (c) and (d), when the dropout probability is greater than 0.5, the performance decreases slightly because too much high dropout will prevent information from spreading between hidden layers. We also noticed that too low or too high lr will cause performance degradation, and the AUC is highest when lr is 0.0005. Therefore, we set dropout to 0.5 and lr to 0.0005 for the VGAE encoder.

\begin{figure}[h]
	\centering
	\centerline{\includegraphics[width=9cm]{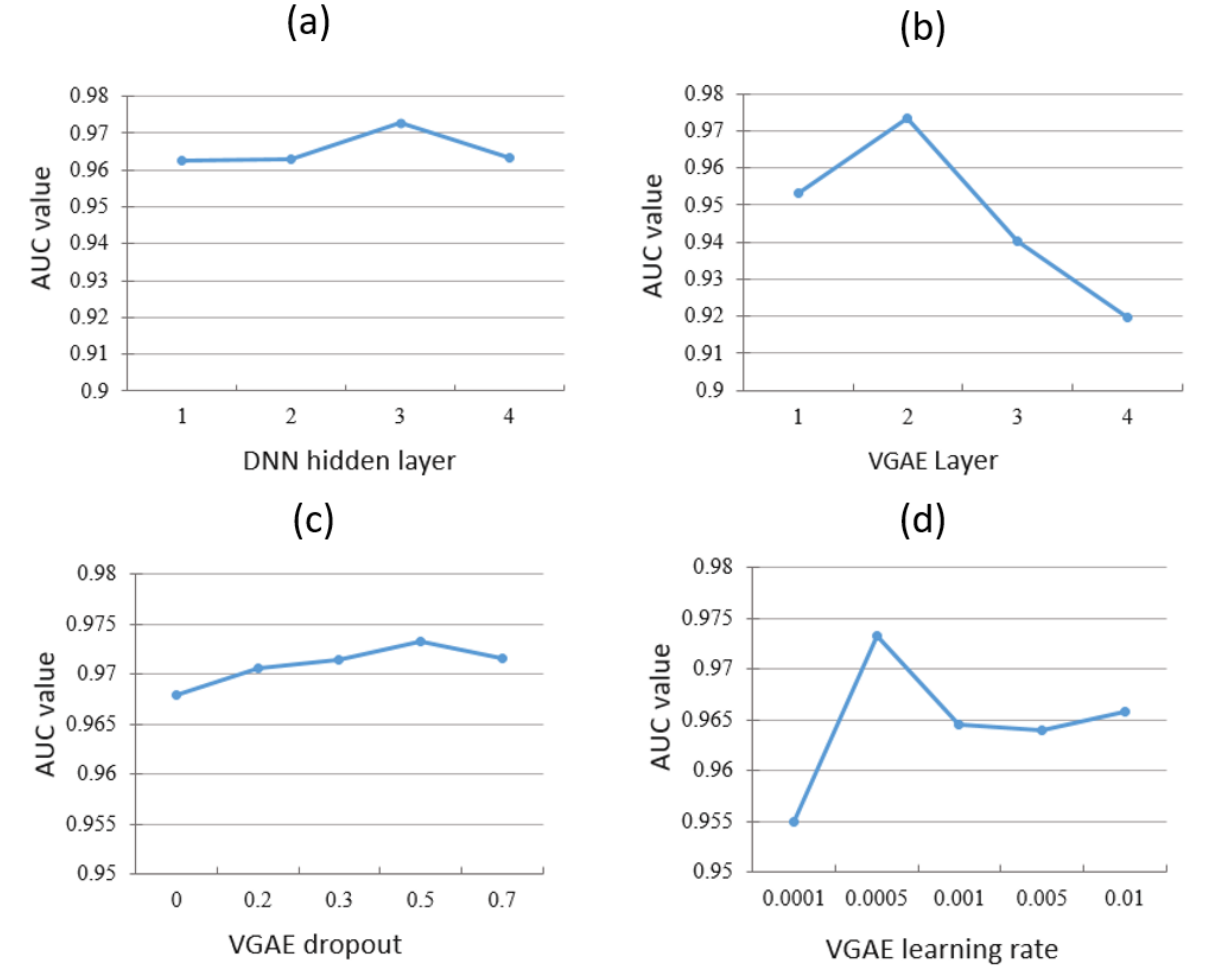}}
	\begin{flushleft}
		\textbf{Fig. 4.} Empirical analysis of the impact of hyperparameters on performance. The subfigures from (a) to (d) show the AUC values corresponding to the number of hidden layers of DNN and VGAE, and the learning rate and loss rate of VGAE, respectively.
	\end{flushleft}
	\label{fig:04}
\end{figure}

\subsection{Performance comparison with SOTA methods}
To benchmark the performance of our proposed method, we compared our method with existing methods proposed for microbe-drug association prediction, as well as a few methods for link prediction problems in the bioinformatics field. We briefly summarized these methods as below:
\begin{itemize}
\item \textbf{EGATMDA} (\citealp{long2020ensembling}) is proposed to predict human microbe-drug associations by constructing a novel integrated graph attention network.
\item \textbf{GCNMDA} (\citealp{long2020predicting}) is a method based on graph convolutional network and conditional random field to predict human microbe-drug association.
\item \textbf{HNERMDA} (\citealp{long2020association}) proposes an embedding representation based on heterogeneous networks for predicting human microbe-drug associations.
\item \textbf{HMDAKATZ} (\citealp{chen2017novel}) is developed specifically for microbial drug prediction based on the KATZ metric.	
\item \textbf{NTSHMDA} (\citealp{luo2018ntshmda}) adopts random walk with restart model to predict microbe-disease associations.
\item \textbf{LAGCN} (\citealp{yu2020predicting}) is a model based on the convolutional neural network with attention mechanism for predicting drug-disease associations.
\end{itemize}

\begin{figure}[h]
	\centering
	\centerline{\includegraphics[width=8cm]{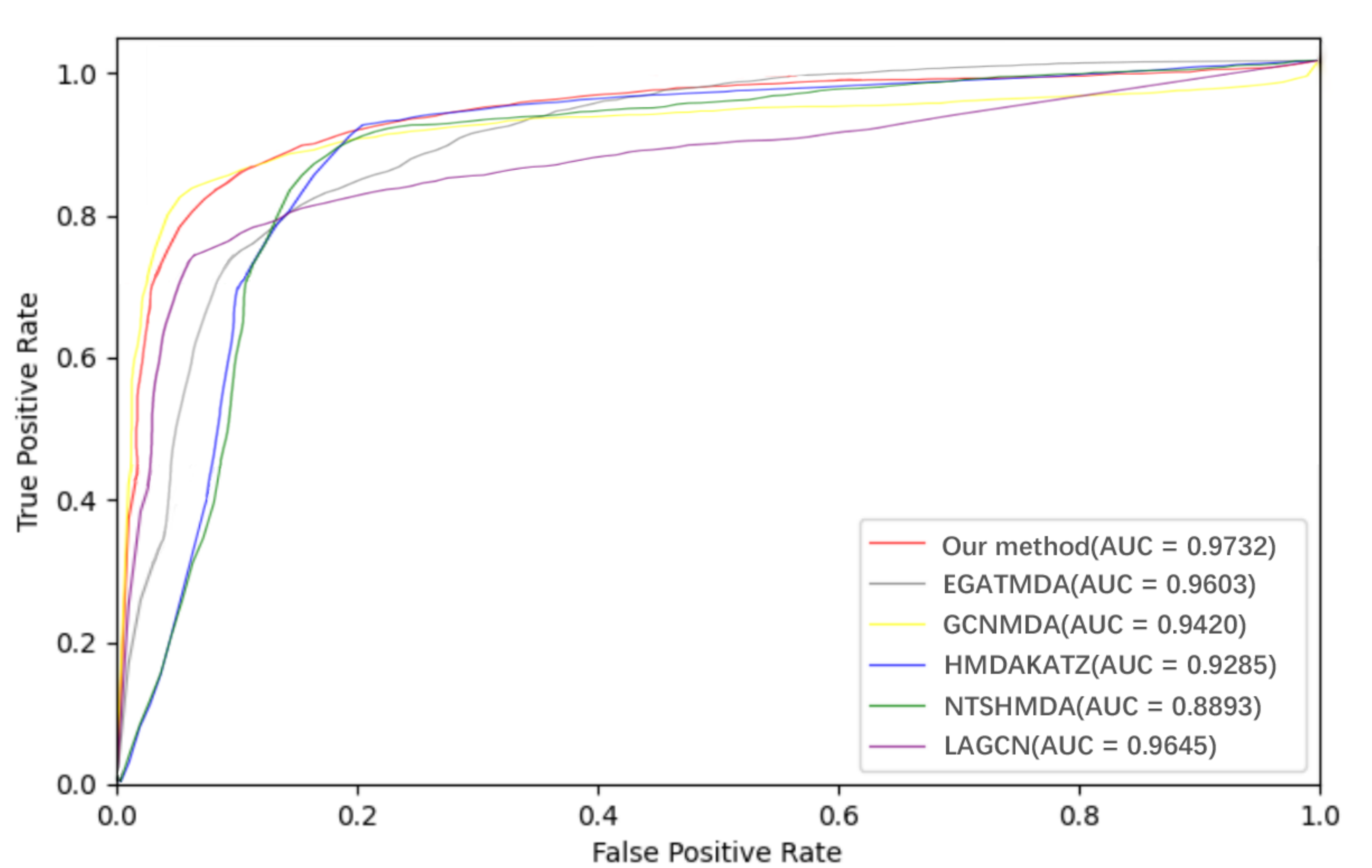}}
	\caption{ROC curves and AUC values of our method and six competitive methods for predicting microbe–drug association on MDAD dataset}
	\label{fig:02}
\end{figure}

We ran these methods on the MDAD dataset with their default parameter settings, and adopt the AUC value as the performance measure. We performed 10-fold cross-validation (CV) for all methods, that is, the experimentally confirmed microbe-drug associations are randomly divided into 10 subsets with equal size, each subset is taken in turn as a test set and the remaining are used to train the model. The average prediction accuracy over the 10-folds is used as the final performance measure. To eliminate the bias of random sampling, the process is repeated 10 times, and the final AUC score was calculated from the average of 10 repetitions. As shown in the figure~\ref{fig:02}, our Graph2MDA model achieved the highest AUC value of 0.9732, followed by LAGCN with 0.9645 AUC value, while NTSHMDA gets the lowest AUC value 0.8893. Overall, the deep learning-based methods outperform those based on traditional machine learning.

\subsection{Model ablation study}
We went further to conduct ablation experiments to evaluate the impact of multi-modal attribute graphs on model performance. First, we tested three different input graphs, including bipartite network, heterogeneous network, and both of them, to perform ablation studies on the MDAD dataset. Specifically, the bipartite network includes only microbe-drug associations, while the heterogeneous network includes microbe-drug, drug-drug, and microbe-microbe associations. Table \ref{tab2} shows the AUC values achieved by our model upon three different input graphs. For comprehensive contrast, we ran both 5-fold and 10-fold cross validations. We found that the best performance could be achieved when both networks are fed into the model, suggesting that the combination of multi-source information facilitates the feature extraction for the prediction of associations between microbes and drugs.

\vspace{-0.3 cm}
\begin{table}[h]
	\centering
	\caption{Performance comparison between different input graph structures \label{tab2}}
	\scalebox{1}{
		\begin{tabular}{m{3.9cm}m{1.8cm}p{2cm}}
			\toprule
			\textbf{Networks} & \textbf{5-flod CV} & \textbf{10-flod CV}       \\ \midrule
			Bipartite network         &0.9477$\pm$0.0015          & 0.9710$\pm$0.0030          \\
			Heterogeneous network      & 0.9190$\pm$0.0045          & 0.9358$\pm$0.0056          \\
			Bipartite + Heterogeneous network        & \textbf{0.9567$\pm$0.0039} & \textbf{0.9732$\pm$0.0037}  \\
			\bottomrule
	\end{tabular}}
\end{table}

We also evaluated the robustness of our model upon multi-modal node attributes. Under the premise of using the BipNet+HetNet network, we independently tested the similarity attribute, secondary attribute, and both of them. Note that when both type attributes are used, the attribute vector is concatenated for each node. The AUC values gained by 5-fold and 10-fold cross validations are shown in Table \ref{tab3}, and the best performance can be achieved when both node attributes are fed into the model. Meanwhile, we found that the secondary attributes also obtain notably high performance alone. We think the reason may lie in that the secondary attributes contain high-order information. 

\vspace{-0.3 cm}
\begin{table}[h]
	\centering
	\caption{Performance comparison between different node attributes \label{tab3}}
	\scalebox{1}{
		\begin{tabular}{m{3.5cm}m{1.8cm}p{2cm}}
			\toprule
			\textbf{Node attributes} & \textbf{5-flod CV} & \textbf{10-flod CV}       \\ \midrule
			similarity attribute         & 0.8940$\pm$0.0204          & 0.9088$\pm$0.0320          \\
			secondary attribute      & 0.9512$\pm$0.0022          & 0.9693$\pm$0.0024          \\
			similarity + secondary attribute       & \textbf{0.9567$\pm$0.0039} & \textbf{0.9732$\pm$ 0.0037}  \\
			\bottomrule
	\end{tabular}}
\end{table}

\subsection{Performance evaluation on two independent datasets}
To verify the generality of our model, we compared our method with the competitive methods on other two independent datasets, aBiofilm (\citealp{2017aBiofilm}), DrugVirus (\citealp{andersen2020discovery}). The aBiofilm database contains 1,720 unique anti-biofilm drugs that target more than 140 microbes. After filtering out the duplicates, we obtained 2,884 microbe-drug associations covering 1,720 drugs and 140 microbes. The DrugVirus database has manually collected associations between 175 drugs and 95 human viruses from the drug database and related publications, there are 933 clinically or experimentally confirmed drug-virus associations. The detail of these two datasets is shown in Table 4.

\begin{figure}[h]
	\centering
	\centerline{\includegraphics[width=7cm]{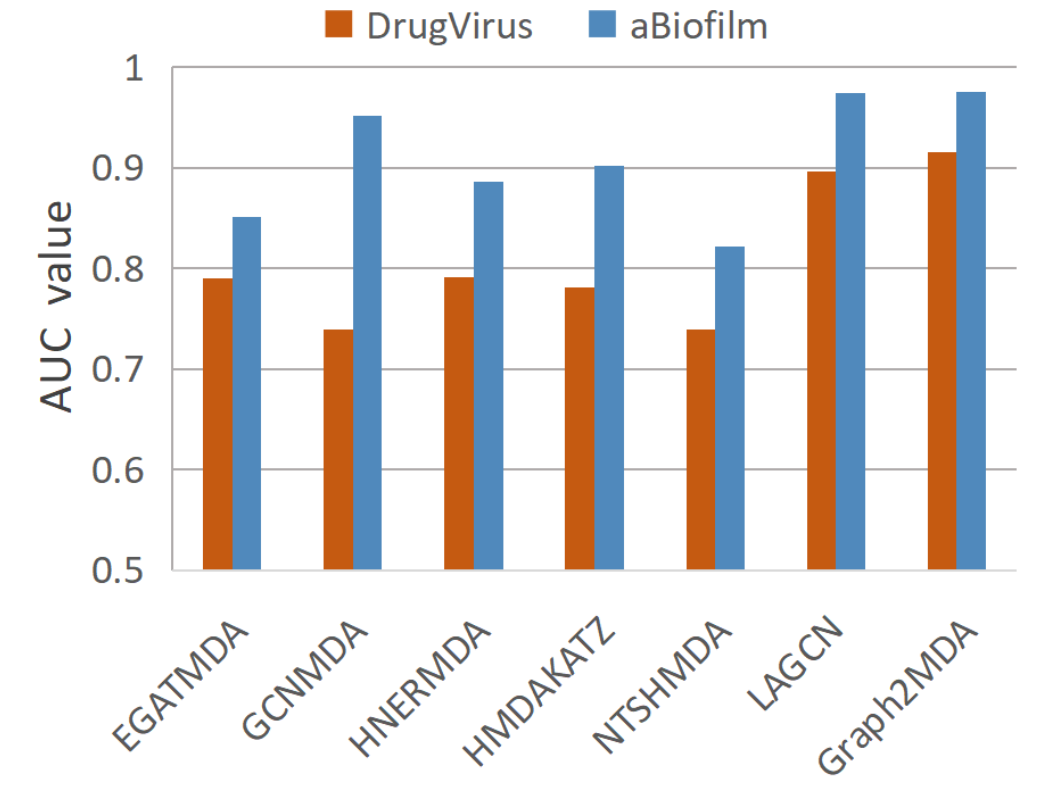}}
	\caption{Performance comparisons of Graph2MDA with six competitive methods on aBiofilm and DrugVrius datasets.}
	\label{fig:03}
\end{figure}

\begin{table}[h]
	\centering
	\caption{Statistics of two independent microbe-drug association dataset \label{tab4}}
	\scalebox{1}{
		\begin{tabular}{p{2cm}m{1.5cm}p{1.5cm}m{1.5cm}}
			\toprule
			\textbf{DataSet} & \textbf{microbes} & \textbf{drugs}    & \textbf{associations}   \\ \midrule
			aBiofilm  & 140 & 1,720  & 2,884    \\
			DrugVirus & 95  & 175   & 933     \\
			\bottomrule
	\end{tabular}}
\end{table}
Similarly, we ran all competitive methods using their default parameter settings on aBiofilm and DrugVirus datasets, and adopt the AUC value as the performance measure. As the DrugVirus dataset is relatively small, we performed a 5-fold cross-validation to avoid too small test set. Figure \ref{fig:03} shows the performance of Graph2MDA and six competitive methods. Compared with other methods, our model still showed the best performance in the two independent datasets. The performance evaluation shows that Graph2MDA is an effective and powerful computational model for predicting the association between microbes and drugs.

\subsection{Interpretation study of latent representation}
\begin{figure}[h]
	\centering
	\centerline{\includegraphics[width=8.5cm]{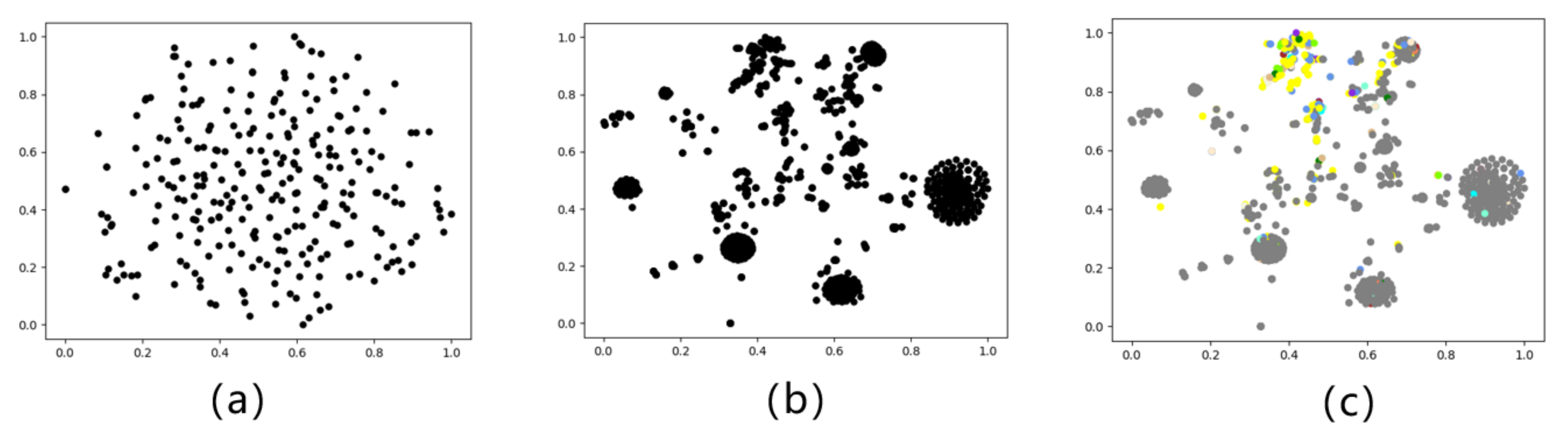}}
	\caption{Scatter plots of drugs in MDAD dataset visualized using the original attributes and learned latent representations by t-SNE tool. (a) Drug distribution visulized in the original attribute space. (b)  Drug distribution visulized in the learned latent representation space. (c) Drugs with mapped ATC codes are colored and others are displayed in gray.}
	\label{fig:tsne}
\end{figure}

Our model can learn high-level feature representations from the original input graph and node attributes. To mine the interpretability, we visualized the learned latent representations of drugs in the MDAD dataset using t-SNE(\citealp{2017Visualizing}), which is a tool to realize high-dimensional data visualization by embedding high-dimensional features into two-dimensional (2D) images. As shown in Figure \ref{fig:tsne} (a), the scatter plot shows the distribution of drugs using the original attributes. It can be seen that before training the drug distribution is in chaos, while with the learned latent representations by our model, the drugs show clear clustering patterns, as shown in Figure \ref{fig:tsne} (b).

To further explore the meaningness of the latent representations, we used the Anatomical Therapeutic Chemical (ATC) code of drugs to verify the consistency between the clustering patterns and ATC classification(\citealp{2009anatomical}). The ATC code is assigned to medicine according to the organ or system it works on and how it works. We manually curated the ATC codes of the 1,373 drugs in MDAD dataset, and successfully map 269 drugs with ATC codes. According to the ATC classification system, drugs belonging to different categories are marked in different colors. For comparison, the drugs without ATC codes are also represented in gray, as shown in Figure 5(c). We found that the yellow scattered spots (yellow corresponding to anti-infectives systemic use) obviously gather together and separate from other clusters. This observation confirms that the learned features of drugs are informative and interpretable, and we make a bold guess that our model can effectively learn the feature that can be translational to drug pharmacological functions.

\subsection{Case studies}
To further verify the effectiveness of Graph2MDA, we applied Graph2MDA to two popular antibacterial drugs, Aloe Vera Gel and Cloxacillin, and two microbes, human immunodeficiency virus (HIV) and mycobacterium tuberculosis as our case studies. For the top 20 predicted microbes or drugs, we cross-checked their synonyms by searching MeSH and DrugBank, and then verify whether the predicted microbe-drug associations have been reported by searching PubMed literature.

Cloxacillin is a semi-synthetic penicillin antibiotic and is widely used to treat $\beta$-hemolytic streptococcal and pneumococcal infections as well as staphylococcal infections. Cloxacillin has an inactivation effect on the penicillinase of most staphylococci and is active against many strains of Staphylococcus aureus and Staphylococcus epidermidis that produce penicillinase (\citealp{2007Cloxacillin}). For the top 20 predicted microbes of cloxacillin, we also found that 15 microbes (75\%)  have been studied, as shown in Table~\ref{Cloxacillin}.  For example, Saengsai \textit{et al.} studied the antimicrobial activity of cloxacillin against the iridoid lactone Plummericin against Enterococcus faecalis and Bacillus subtilis (\citealp{2015Antibacterial}). Orogade and Akuse also reported that cloxacillin can inhibit most to 50\% activity of Staphylococcus aureus, Schistosoma, and Salmonella Typhi.

The drug Aloe Vera Gel has long been used as a traditional medicine to induce wound healing. It is a natural product and is now used in the cosmetics industry. Although there are multiple instructions for use, the benefits of aloe vera are attributed to the polysaccharides contained in the leaf gel (\citealp{gupta2012pharmacological}). Kaithwas et al. also studied the use of aloe vera in constipation, inflammation, cancer, ulcers, and diabetes (\citealp{kaithwas2014evaluation}). As a result, out of the 20 microbes predicted to be targeted by Aloe Vera Gel, 16 microbes (80\%) have been reported by previous studies. As shown in Table~S1, we list the name of microbes and the PMIDs of the publications that reported the associations between the microbes and Aloe Vera Gel. For example, Fani M \textit{et al.} investigated the inhibitory activity of aloe vera gel against certain cariogenic bacteria (Streptococcus mutans), periodontal bacteria (Aggregatibacter actinomycetemcomitans, porphyromonas gingivalis), and opportunistic periodontal bacteria (Bacteroides fragilis) isolated from patients with dental caries and periodontal disease(\citealp{2012Inhibitory})

\begin{table}[h]
	\renewcommand\arraystretch{1}
	\centering
	\caption{The 20 predicted Cloxacillin-related microbes and related publications}
	 \label{Cloxacillin}
	\scalebox{1}{
		\begin{tabular}{m{2cm}m{1.8cm}m{1.6cm}m{1.8cm}}
			\toprule
			\textbf{Microbe} & \textbf{Evidence} & \textbf{Microbe} & \textbf{Evidence}    \\
			\midrule
			Bacillus cereusereus     & PMID:24876650     & Helicobacter pylori   & PMID: 10748053    \\
			Bacillus subtilis        & PMID:25945113     & Klebsiella planticola & Unconfirmed       \\
			Baker's yeast            & PMID:2089228      & Klebsiella pneumoniae & PMID:20597925     \\
			Burkholderia cepacia     & Unconfirmed       & Micrococcus luteus    & PMID:7771695      \\
			Candida albicans         & PMID:2713774      & Pantoea agglomerans   & PMID:33666040     \\
			Candida dubliniensis     & PMID:16353125     & Salmonella Typhi      & PMID:15490798     \\
			Candida spp.             & PMID:21496537     & Schistosoma           & PMID:15490798     \\
			Clostridium pasteurianum & Unconfirmed       & Staphylococcus aureus & PMID:15490798     \\
			Enterobacter aerogenes   & PMID:22001269     & Streptomyces sp.      & PMID:6970744      \\
			Francisella novicida     & Unconfirmed       & Thermus thermophilus  & Unconfirmed       \\
			\bottomrule
	\end{tabular}}
\end{table}

Mycobacterium tuberculosis is a gram-positive aerobic bacteria. This bacterium can invade all organs of the body, but it is the most common cause of tuberculosis. Among the top 20 drugs predicted to target mycobacterium tuberculosis, 80\% of drugs are supported by the literature. Table \ref{mycobacterium} shows the drug names and PMIDs of publications. For example, in PMID29311078, the absolute concentration method showed that the activity of Amikacin against Mycobacterium tuberculosis was higher than kanamycin and capreomycin~(\citealp{2018In}). Apramycin is a unique aminoglycoside with antibacterial activity and ototoxicity. In two murine models of infection, Apramycin has shown significant antibacterial efficacy to Mycobacterium tuberculosis and Staphylococcus aureus\citep{apramycin2014}. The human immunodeficiency virus (HIV) is a virus that can attack the human immune system. HIV will induce the failure and functional damage of key T cells, which will lead to dysfunction and defects of the entire immune system, and ultimately lead to opportunistic infections and tumors (\citealp{1988}). We checked the top 20 predicted drugs and found 95\% of them have been already confirmed, as shown in Table~S2. Take Erythromycin as an example, it has been reported to inhibit HIV-1 replication in macrophages through modulation of MAPK activity to induce small isoforms of C/EBP$\beta$ (\citealp{2008Erythromycin}).

\begin{table}[h]
	\renewcommand\arraystretch{1}
	\centering
	\caption{Top 20 predicted mycobacterium tuberculosis-related drugs and related publications. \label{mycobacterium}}
	\scalebox{1}{
		\begin{tabular}{m{1.6cm}m{2cm}m{1.6cm}m{1.8cm}}
			\toprule
			\textbf{Microbe} & \textbf{Evidence} & \textbf{Microbe} & \textbf{Evidence}    \\
			\midrule
		Amikacin         & PMID:29311078     & Gatifloxacin     & PMID:16714850     \\
		Amprenavir       & Unconfirmed       & Genistein        & Unconfirmed       \\
		Apramycin        & PMID:25136009     & Gentamicin       & PMID:22143521     \\
		Azithromycin     & PMID:7849341      & Indinavir        & PMID:21442799     \\
		Calanolide A     & PMID:14980631     & Lopinavir        & PMID:21442799     \\
		Cloxacillin      & PMID: 25104892    & Minocycline      & PMID:30597040     \\
		Darunavir        & PMID:28193650     & Nevirapine       & PMID:2039216      \\
		Dirithromycin    & Unconfirmed       & Raltegravir      & PMID:30350998     \\
		Dolutegravir     & PMID:33315751     & Rilpivirine      & Unconfirmed       \\
		Desipramine      & PMID:7649718      & inapic acid      & Unconfirmed     \\
			\bottomrule
	\end{tabular}}
\end{table}

\section{Conclusion}

In this paper, we propose a novel model based on Graph2MDA, an integrated framework of Variational Graph Autoencoder(VGAE) and Deep Neural Networks (DNN), to predict the correlation between microbes and drugs. The experimental results show that our proposed Graph2MDA model is better than the existing state-of-the-art methods. We think the main contributions of our work lie in at least three aspects: 1) We constructed  multi-modal attribute graphs to effectively integrate the rich similarity and ontology information of microbes and drugs. 2) A novel framework based on VGAE and DNN was proposed to predict the association between microbes and drugs. Compared with current state-of-the-art methods for predicting microbe-drug association based on graph neural networks and link prediction, our method achieves better performance than competitive methods on three independent datasets. 3) We verified that the learned latent representations are sematically related to drug pharmacological functions. Specifically, we found the drugs show obvious clustering patterns in the latent representation space, and the clusters are significantly consistent with drug ATC classification.


\section*{Funding}
This work was supported by the National Natural Science Foundation of China under grants No.~61972422 and No.~62072058.

\noindent
\emph{Conflict of interest}: none declared.

\bibliographystyle{plainnat}
\bibliography{document}

\begin{thebibliography}{46}
\providecommand{\natexlab}[1]{#1}
\providecommand{\url}[1]{\texttt{#1}}
\expandafter\ifx\csname urlstyle\endcsname\relax
  \providecommand{\doi}[1]{doi: #1}\else
  \providecommand{\doi}{doi: \begingroup \urlstyle{rm}\Url}\fi

\bibitem[Andersen et~al.(2020)Andersen, Ianevski, Lysvand, Vitkauskiene,
  Oksenych, Bjørås, Telling, Lutsar, Dumpis, Irie, Tenson, Kantele, and
  Kainov]{andersen2020discovery}
P.~I. Andersen, A.~Ianevski, H.~Lysvand, A.~Vitkauskiene, V.~Oksenych,
  M.~Bjørås, K.~Telling, I.~Lutsar, U.~Dumpis, Y.~Irie, T.~Tenson,
  A.~Kantele, and D.~E. Kainov.
\newblock {{D}iscovery and development of safe-in-man broad-spectrum antiviral
  agents}.
\newblock \emph{Int J Infect Dis}, 93:\penalty0 268--276, Apr 2020.

\bibitem[Castle(2007)]{2007Cloxacillin}
S.~S. Castle.
\newblock Cloxacillin - sciencedirect.
\newblock \emph{xPharm: The Comprehensive Pharmacology Reference}, 305\penalty0
  (7900):\penalty0 1--5, 2007.

\bibitem[Chen et~al.(2018)Chen, Huang, You, Yan, and Wang]{chen2017novel}
X.~Chen, Y.~A. Huang, Z.~H. You, G.~Y. Yan, and X.~S. Wang.
\newblock {{A} novel approach based on {K}{A}{T}{Z} measure to predict
  associations of human microbiota with non-infectious diseases}.
\newblock \emph{Bioinformatics}, 34\penalty0 (8):\penalty0 1440, 04 2018.

\bibitem[Dijkstra et~al.(2018)Dijkstra, van~der Laan, Akkerman, Bolhuis,
  de~Lange, and Kosterink]{2018In}
J.~A. Dijkstra, T.~van~der Laan, O.~W. Akkerman, M.~S. Bolhuis, W.~C.~M.
  de~Lange, and J.~G.~W. Kosterink.
\newblock {{S}usceptibility of {M}ycobacterium tuberculosis to {A}mikacin,
  {K}anamycin, and {C}apreomycin}.
\newblock \emph{Antimicrob Agents Chemother}, 62\penalty0 (3), 03 2018.

\bibitem[Ding et~al.(2021)Ding, Tian, Lei, Liao, and Wu]{ding2020variational}
Y.~Ding, L.~P. Tian, X.~Lei, B.~Liao, and F.~X. Wu.
\newblock {{V}ariational graph auto-encoders for mi{R}{N}{A}-disease
  association prediction}.
\newblock \emph{Methods}, 192:\penalty0 25--34, 08 2021.

\bibitem[Fani and Kohanteb(2012)]{2012Inhibitory}
M.~Fani and J.~Kohanteb.
\newblock {{I}nhibitory activity of {A}loe vera gel on some clinically isolated
  cariogenic and periodontopathic bacteria}.
\newblock \emph{J Oral Sci}, 54\penalty0 (1):\penalty0 15--21, Mar 2012.

\bibitem[Gupta and Malhotra(2012)]{gupta2012pharmacological}
V.~K. Gupta and S.~Malhotra.
\newblock {{P}harmacological attribute of {A}loe vera: {R}evalidation through
  experimental and clinical studies}.
\newblock \emph{Ayu}, 33\penalty0 (2):\penalty0 193--196, Apr 2012.

\bibitem[Haiser et~al.(2013)Haiser, Gootenberg, Chatman, Sirasani, Balskus, and
  Turnbaugh]{haiser2013predicting}
H.~J. Haiser, D.~B. Gootenberg, K.~Chatman, G.~Sirasani, E.~P. Balskus, and
  P.~J. Turnbaugh.
\newblock {{P}redicting and manipulating cardiac drug inactivation by the human
  gut bacterium {E}ggerthella lenta}.
\newblock \emph{Science}, 341\penalty0 (6143):\penalty0 295--298, Jul 2013.

\bibitem[Hattori et~al.(2010)Hattori, Tanaka, Kanehisa, and
  Goto]{hattori2010simcomp}
M.~Hattori, N.~Tanaka, M.~Kanehisa, and S.~Goto.
\newblock {{S}{I}{M}{C}{O}{M}{P}/{S}{U}{B}{C}{O}{M}{P}: chemical structure
  search servers for network analyses}.
\newblock \emph{Nucleic Acids Res}, 38\penalty0 (Web Server issue):\penalty0
  W652--656, Jul 2010.

\bibitem[He et~al.(2016)He, Zhang, Ren, and Sun]{he2016deep}
K.~He, X.~Zhang, S.~Ren, and J.~Sun.
\newblock Deep residual learning for image recognition.
\newblock 2016.

\bibitem[Hinton(2017)]{2017Visualizing}
Pk~Geoffrey Hinton.
\newblock Visualizing data using t-sne laurens van der maaten micc-ikat.
\newblock 2017.

\bibitem[Huttenhower et~al.(2012)Huttenhower, Gevers, and Knight]{0Structure}
C.~Huttenhower, D.~Gevers, and R.~Knight.
\newblock {{S}tructure, function and diversity of the healthy human
  microbiome}.
\newblock \emph{Nature}, 486\penalty0 (7402):\penalty0 207--214, Jun 2012.

\bibitem[Jain et~al.(2018)Jain, Zhang, and Huang]{jain2018random}
D.~K. Jain, Z.~Zhang, and K.~Huang.
\newblock Random walk-based feature learning for micro-expression recognition.
\newblock \emph{Pattern Recognition Letters}, 115\penalty0 (NOV.1):\penalty0
  92--100, 2018.

\bibitem[Kaithwas et~al.(2014)Kaithwas, Singh, and
  Bhatia]{kaithwas2014evaluation}
G.~Kaithwas, P.~Singh, and D.~Bhatia.
\newblock {{E}valuation of in vitro and in vivo antioxidant potential of
  polysaccharides from {A}loe vera ({A}loe barbadensis {M}iller) gel}.
\newblock \emph{Drug Chem Toxicol}, 37\penalty0 (2):\penalty0 135--143, Apr
  2014.

\bibitem[Kamneva(2017)]{kamneva2017genome}
O.~K. Kamneva.
\newblock {{G}enome composition and phylogeny of microbes predict their
  co-occurrence in the environment}.
\newblock \emph{PLoS Comput Biol}, 13\penalty0 (2):\penalty0 e1005366, 02 2017.

\bibitem[Kashyap et~al.(2017)Kashyap, Chia, Nelson, Segal, and
  Elinav]{2017Microbiome}
P.~C. Kashyap, N.~Chia, H.~Nelson, E.~Segal, and E.~Elinav.
\newblock {{M}icrobiome at the {F}rontier of {P}ersonalized {M}edicine}.
\newblock \emph{Mayo Clin Proc}, 92\penalty0 (12):\penalty0 1855--1864, Dec
  2017.

\bibitem[Kingma and Welling(2014)]{kingma2013auto}
D.~P. Kingma and M.~Welling.
\newblock Auto-encoding variational bayes.
\newblock 2014.

\bibitem[Kip and Welling(2016)]{kipf2016semi}
T.~N. Kip and M.~Welling.
\newblock Semi-supervised classification with graph convolutional networks.
\newblock 2016.

\bibitem[Kipf and Welling(2016)]{kipf2016variational}
T.~N. Kipf and M.~Welling.
\newblock Variational graph auto-encoders.
\newblock 2016.

\bibitem[Klatt et~al.(2017)Klatt, Cheu, Birse, Zevin, and Perner]{2017Vaginal}
N.~R. Klatt, R.~Cheu, K.~Birse, A.~S. Zevin, and M.~Perner.
\newblock {{V}aginal bacteria modify {H}{I}{V} tenofovir microbicide efficacy
  in {A}frican women}.
\newblock \emph{Science}, 356\penalty0 (6341):\penalty0 938--945, 06 2017.

\bibitem[Komuro et~al.(2008)Komuro, Sunazuka, Akagawa, Yokota, Iwamoto, and
  Omura]{2008Erythromycin}
I.~Komuro, T.~Sunazuka, K.~S. Akagawa, Y.~Yokota, A.~Iwamoto, and S.~Omura.
\newblock {{E}rythromycin derivatives inhibit {H}{I}{V}-1 replication in
  macrophages through modulation of {M}{A}{P}{K} activity to induce small
  isoforms of {C}/{E}{B}{P}beta}.
\newblock \emph{Proc Natl Acad Sci U S A}, 105\penalty0 (34):\penalty0
  12509--12514, Aug 2008.

\bibitem[Long and Luo(2021)]{long2020association}
Y.~Long and J.~Luo.
\newblock {{A}ssociation {M}ining to {I}dentify {M}icrobe {D}rug {I}nteractions
  {B}ased on {H}eterogeneous {N}etwork {E}mbedding {R}epresentation}.
\newblock \emph{IEEE J Biomed Health Inform}, 25\penalty0 (1):\penalty0
  266--275, 01 2021.

\bibitem[Long et~al.(2020{\natexlab{a}})Long, Wu, Kwoh, Luo, and
  Li]{long2020predicting}
Y.~Long, M.~Wu, C.~K. Kwoh, J.~Luo, and X.~Li.
\newblock {{P}redicting human microbe-drug associations via graph convolutional
  network with conditional random field}.
\newblock \emph{Bioinformatics}, 36\penalty0 (19):\penalty0 4918--4927, 12
  2020{\natexlab{a}}.

\bibitem[Long et~al.(2020{\natexlab{b}})Long, Wu, Liu, Kwoh, Luo, and
  Li]{long2020ensembling}
Y.~Long, M.~Wu, Y.~Liu, C.~K. Kwoh, J.~Luo, and X.~Li.
\newblock {{E}nsembling graph attention networks for human microbe-drug
  association prediction}.
\newblock \emph{Bioinformatics}, 36\penalty0 (Suppl-2):\penalty0 i779--i786, 12
  2020{\natexlab{b}}.

\bibitem[Luo and Long(2020)]{luo2018ntshmda}
J.~Luo and Y.~Long.
\newblock {{N}{T}{S}{H}{M}{D}{A}: {P}rediction of {H}uman {M}icrobe-{D}isease
  {A}ssociation {B}ased on {R}andom {W}alk by {I}ntegrating {N}etwork
  {T}opological {S}imilarity}.
\newblock \emph{IEEE/ACM Trans Comput Biol Bioinform}, 17\penalty0
  (4):\penalty0 1341--1351, 2020.

\bibitem[Macpherson and Harris(2004)]{2004Interactions}
A.~J. Macpherson and N.~L. Harris.
\newblock {{I}nteractions between commensal intestinal bacteria and the immune
  system}.
\newblock \emph{Nat Rev Immunol}, 4\penalty0 (6):\penalty0 478--485, Jun 2004.

\bibitem[Malla et~al.(2018)Malla, Dubey, Kumar, Yadav, Hashem, and
  Abd~Allah]{2018Exploring}
M.~A. Malla, A.~Dubey, A.~Kumar, S.~Yadav, A.~Hashem, and E.~F. Abd~Allah.
\newblock {{E}xploring the {H}uman {M}icrobiome: {T}he {P}otential {F}uture
  {R}ole of {N}ext-{G}eneration {S}equencing in {D}isease {D}iagnosis and
  {T}reatment}.
\newblock \emph{Front Immunol}, 9:\penalty0 2868, 2018.

\bibitem[Meng et~al.(2021)Meng, Jin, Tang, and Xu]{meng2021drug}
Y.~Meng, M.~Jin, X.~Tang, and J.~Xu.
\newblock {{D}rug repositioning based on similarity constrained probabilistic
  matrix factorization: {C}{O}{V}{I}{D}-19 as a case study}.
\newblock \emph{Appl Soft Comput}, 103:\penalty0 107135, May 2021.

\bibitem[Meyer et~al.(2014)Meyer, Freihofer, Scherman, Teague, Lenaerts, and
  Böttger]{apramycin2014}
M.~Meyer, P.~Freihofer, M.~Scherman, J.~Teague, A.~Lenaerts, and E.~C.
  Böttger.
\newblock {{I}n vivo efficacy of apramycin in murine infection models}.
\newblock \emph{Antimicrob Agents Chemother}, 58\penalty0 (11):\penalty0
  6938--6941, Nov 2014.

\bibitem[Nahler(2009)]{2009anatomical}
G.~Nahler.
\newblock anatomical therapeutic chemical classification system (atc).
\newblock \emph{Springer Vienna}, 2009.

\bibitem[Rajput et~al.(2018)Rajput, Thakur, Sharma, and Kumar]{2017aBiofilm}
A.~Rajput, A.~Thakur, S.~Sharma, and M.~Kumar.
\newblock {a{B}iofilm: a resource of anti-biofilm agents and their potential
  implications in targeting antibiotic drug resistance}.
\newblock \emph{Nucleic Acids Res}, 46\penalty0 (D1):\penalty0 D894--D900, 01
  2018.

\bibitem[Saengsai et~al.(2015)Saengsai, Kongtunjanphuk, Yoswatthana, Kummalue,
  and Jiratchariyakul]{2015Antibacterial}
J.~Saengsai, S.~Kongtunjanphuk, N.~Yoswatthana, T.~Kummalue, and
  W.~Jiratchariyakul.
\newblock {{A}ntibacterial and {A}ntiproliferative {A}ctivities of
  {P}lumericin, an {I}ridoid {I}solated from {M}omordica charantia {V}ine}.
\newblock \emph{Evid Based Complement Alternat Med}, 2015:\penalty0 823178,
  2015.

\bibitem[Schwabe and Jobin(2013)]{2013cancer}
R.~F. Schwabe and C.~Jobin.
\newblock The microbiome and cancer.
\newblock \emph{Nature Reviews Cancer}, 13\penalty0 (11):\penalty0 800--812,
  2013.

\bibitem[Shi et~al.(2021)Shi, Zhang, Jin, Quan, and Yin]{shi2021representation}
Z.~Shi, H.~Zhang, C.~Jin, X.~Quan, and Y.~Yin.
\newblock {{A} representation learning model based on variational inference and
  graph autoencoder for predicting lnc{R}{N}{A}-disease associations}.
\newblock \emph{BMC Bioinformatics}, 22\penalty0 (1):\penalty0 136, Mar 2021.

\bibitem[Sommer and Backhed(2013)]{2013The}
F.~Sommer and F.~Backhed.
\newblock {{T}he gut microbiota--masters of host development and physiology}.
\newblock \emph{Nat Rev Microbiol}, 11\penalty0 (4):\penalty0 227--238, Apr
  2013.

\bibitem[Stingl(1988)]{1988}
G.~Stingl.
\newblock {[{H}{I}{V}-1 infection: pathogenesis of immune suppression]}.
\newblock \emph{Wien Med Wochenschr}, 138\penalty0 (19-20):\penalty0 487--492,
  Oct 1988.

\bibitem[Sun et~al.(2018)Sun, Zhang, Cai, Ming, Li, and Chen]{sun2018mdad}
Y.~Z. Sun, D.~H. Zhang, S.~B. Cai, Z.~Ming, J.~Q. Li, and X.~Chen.
\newblock {{M}{D}{A}{D}: {A} {S}pecial {R}esource for {M}icrobe-{D}rug
  {A}ssociations}.
\newblock \emph{Front Cell Infect Microbiol}, 8:\penalty0 424, 2018.

\bibitem[Sutradhar et~al.(2021)Sutradhar, Ching, Desai, Suprenant, Briars,
  Heins, Khalil, and Zaman]{2014Antimicrobial}
I.~Sutradhar, C.~Ching, D.~Desai, M.~Suprenant, E.~Briars, Z.~Heins, A.~S.
  Khalil, and M.~H. Zaman.
\newblock {{C}omputational {M}odel {T}o {Q}uantify the {G}rowth of
  {A}ntibiotic-{R}esistant {B}acteria in {W}astewater}.
\newblock \emph{mSystems}, 6\penalty0 (3):\penalty0 e0036021, Jun 2021.

\bibitem[Szklarczyk et~al.(2021)Szklarczyk, Gable, Nastou, Lyon, and
  Kirsch]{2020The}
D.~Szklarczyk, A.~L. Gable, K.~C. Nastou, D.~Lyon, and R.~Kirsch.
\newblock {{T}he {S}{T}{R}{I}{N}{G} database in 2021: customizable
  protein-protein networks, and functional characterization of user-uploaded
  gene/measurement sets}.
\newblock \emph{Nucleic Acids Res}, 49\penalty0 (D1):\penalty0 D605--D612, 01
  2021.

\bibitem[Ventura et~al.(2009)Ventura, O'Flaherty, Claesson, Turroni,
  Klaenhammer, van Sinderen, and O'Toole]{2009Genome}
M.~Ventura, S.~O'Flaherty, M.~J. Claesson, F.~Turroni, T.~R. Klaenhammer,
  D.~van Sinderen, and P.~W. O'Toole.
\newblock {{G}enome-scale analyses of health-promoting bacteria:
  probiogenomics}.
\newblock \emph{Nat Rev Microbiol}, 7\penalty0 (1):\penalty0 61--71, Jan 2009.

\bibitem[Weber et~al.(2016)Weber, Knebel, Strassburger, Kotzka, Stehle,
  Szendroedi, Müssig, Buyken, and Roden]{chen2002principle}
K.~S. Weber, B.~Knebel, K.~Strassburger, J.~Kotzka, P.~Stehle, J.~Szendroedi,
  K.~Müssig, A.~E. Buyken, and M.~Roden.
\newblock {{A}ssociations between explorative dietary patterns and serum lipid
  levels and their interactions with {A}po{A}5 and {A}po{E} haplotype in
  patients with recently diagnosed type 2 diabetes}.
\newblock volume~15, page 138, Sep 2016.

\bibitem[Wen et~al.(2008)Wen, Ley, Volchkov, Stranges, Avanesyan, Stonebraker,
  Hu, Wong, Szot, Bluestone, Gordon, and Chervonsky]{2008Innate}
L.~Wen, R.~E. Ley, P.~Y. Volchkov, P.~B. Stranges, L.~Avanesyan, A.~C.
  Stonebraker, C.~Hu, F.~S. Wong, G.~L. Szot, J.~A. Bluestone, J.~I. Gordon,
  and A.~V. Chervonsky.
\newblock {{I}nnate immunity and intestinal microbiota in the development of
  {T}ype 1 diabetes}.
\newblock \emph{Nature}, 455\penalty0 (7216):\penalty0 1109--1113, Oct 2008.

\bibitem[Wishart et~al.(2018)Wishart, Feunang, Guo, Lo, Marcu, Grant, and
  Sajed]{wishart2018drugbank}
D.~S. Wishart, Y.~D. Feunang, A.~C. Guo, E.~J. Lo, A.~Marcu, J.~R. Grant, and
  T.~Sajed.
\newblock {{D}rug{B}ank 5.0: a major update to the {D}rug{B}ank database for
  2018}.
\newblock \emph{Nucleic Acids Res}, 46\penalty0 (D1):\penalty0 D1074--D1082, 01
  2018.

\bibitem[Yu et~al.(2021)Yu, Huang, Zhao, Xiao, and Zhang]{yu2020predicting}
Z.~Yu, F.~Huang, X.~Zhao, W.~Xiao, and W.~Zhang.
\newblock {{P}redicting drug-disease associations through layer attention graph
  convolutional network}.
\newblock \emph{Brief Bioinform}, 22\penalty0 (4), Jul 2021.

\bibitem[Zhang et~al.(2009)Zhang, DiBaise, Zuccolo, Kudrna, Braidotti, Yu,
  Parameswaran, Crowell, Wing, Rittmann, and Krajmalnik-Brown]{2010Human}
H.~Zhang, J.~K. DiBaise, A.~Zuccolo, D.~Kudrna, M.~Braidotti, Y.~Yu,
  P.~Parameswaran, M.~D. Crowell, R.~Wing, B.~E. Rittmann, and
  R.~Krajmalnik-Brown.
\newblock {{H}uman gut microbiota in obesity and after gastric bypass}.
\newblock \emph{Proc Natl Acad Sci U S A}, 106\penalty0 (7):\penalty0
  2365--2370, Feb 2009.

\bibitem[Zimmermann et~al.(2021)Zimmermann, Patil, Typas, and
  Maier]{2021Towards}
M.~Zimmermann, K.~R. Patil, A.~Typas, and L.~Maier.
\newblock {{T}owards a mechanistic understanding of reciprocal drug-microbiome
  interactions}.
\newblock \emph{Mol Syst Biol}, 17\penalty0 (3):\penalty0 e10116, 03 2021.

\end{thebibliography}

\end{document}